\documentclass[pre,reprint,nobibnotes, aps,showpacs]{revtex4-1}
\usepackage[english]{babel}
\usepackage{grffile}
\usepackage{graphicx,epstopdf,color}
\usepackage{amsmath,amssymb,psfrag,mathtools}
\usepackage[caption=false]{subfig}

\usepackage[draft]{fixme}
\usepackage[T1]{fontenc}
\usepackage[latin1]{inputenc}

\newcommand{\Pout}{P_{\text{out}}}
\newcommand{\Poutz}{P_{\text{out},0}}
\newcommand{\Pin}{P_{\mathrm{in}}}

\newcommand{\rhold}{\varrho_{\text{LD}}}
\newcommand{\rhohd}{\varrho_{\text{HD}}}
\newcommand{\rhomc}{\varrho_{\text{MC}}}

\newcommand{\jc}{J^{\text{in}}}
\newcommand{\jsc}{j^{\text{in}}}
\newcommand{\jm}{J^{\text{out}}}
\newcommand{\jsm}{j^{\text{out}}}

\newcommand{\win}{w_{\text{in}}}
\newcommand{\wout}{w_{\text{out}}}
\newcommand{\woutz}{w_{\text{out},0}}
\newcommand{\rhovb}{\bar{\rho}_v}
\newcommand{\rhob}{\bar{\rho}}

\DeclarePairedDelimiter{\ceil}{\lceil}{\rceil}
\DeclareMathOperator*{\argmax}{\arg\!\max}
\DeclareMathOperator*{\argmin}{\arg\!\min}

\begin{document}

\title{Efficiency at maximum power of motor traffic on networks}
\author{N. Golubeva}
\author{A. Imparato}
\affiliation{Department of Physics and Astronomy, University of Aarhus, Ny Munkegade, Building 1520, DK--8000 Aarhus C, Denmark}
\date{\today}

\begin{abstract}
We study motor traffic on Bethe networks subject to hard-core exclusion for both tightly coupled one-state machines and loosely coupled two-state machines that perform work against a constant load. In both cases we find an interaction-induced enhancement of the efficiency at maximum power (EMP) as compared to non-interacting motors. The EMP enhancement occurs for a wide range of network and single motor parameters and is due to a change in the characteristic load-velocity relation caused by phase transitions in the system. Using a quantitative measure of the trade-off between the EMP enhancement and the corresponding loss in the maximum output power we identify parameter regimes where motor traffic systems operate efficiently at maximum power without a significant decrease in the maximum power output due to jamming effects. 
\end{abstract}

\maketitle

\section{Introduction} 
Molecular motors are nanosized biological machines involved in essential cellular processes such as intracellular transport, protein synthesis and transcription and repair of DNA \cite{Alberts2007}. Over the last two decades molecular machines have been studied extensively both theoretically and experimentally in order to reveal the organizing principles behind their ability to efficiently solve specialized tasks in complex environments (see \cite{Seifert2011,Kolomeisky2007,Veigel2011a} and references therein). More recently, there has been a rapid development of various artificial nanomotors with the aim of mimicking the performance of biological machines \cite{Kay2007,Liu2009,Lund2010}. In a typical setup the nanomachine translates along a track in an isothermal environment driven by chemical reactions, external forces, electric or magnetic fields. While an external modification of the system parameters is often needed to operate the system, the goal is to design autonomous machines that operate under steady-state conditions in anology to their biological counterparts \cite{Liu2009,Lund2010}. 

Since motor proteins often function collectively in the cell, a detailed understanding of molecular motor function requires considering the role of cooperative effects mediated through, e.g., excluded volume interactions or mechanical constraints imposed by motors being coupled to the same cargo \cite{Holzbaur2010,Guerin2010}. Likewise, man-made molecular motors must operate in unison in order to achieve the desired efficiency and fidelity \cite{Rank2013}. The collective motion of molecular motors on an underlying substrate is frequently referred to as (molecular) motor traffic.  

The efficiency of isothermal machines defined as the delivered power output divided by the consumed power input is constrained by the thermodynamic bound 1. However, achieving maximum efficiency comes at the expense of quasistatic operation and, hence, zero power output. A practically more relevant quantity to consider is thus the efficiency at maximum power (EMP) which offers a quantitative measure of the power-efficiency trade-off in nanomotors. Both the universal and system-specific features of the EMP for single isothermal machines have been addressed in \cite{Golubeva2012,Seifert2011a,VandenBroeck2012}, while the issue of the EMP for cooperative systems has received far less attention to date. Motor traffic is a widely studied phenomenon, which is typically modelled using exclusion processes on discrete lattices \cite{ASEP,Golubeva2012a,Golubeva2013,Lipowsky2001,Nishinari2005,Garai2009,Tripathi2009,Klumpp2008a,Ciandrini2010,Neri2011,Neri2013a,Neri2013}. Even so, most of these works are concerned with collective dynamics rather than thermodynamics, and they often deal with specific molecular motor proteins \cite{Lipowsky2001,Nishinari2005,Garai2009,Tripathi2009,Klumpp2008a,Ciandrini2010}. Our previous papers \cite{Golubeva2012a,Golubeva2013} in which we studied the effect of exclusion interactions on the thermodynamic aspects of intracellular traffic, pose an exception. In these works we found that the EMP is enhanced, as compared to the non-interacting case, due to an interaction-induced change in the characteristic response of the system to external loads. Yet, this result was obtained in the specific context of kinesin motors moving on a single filament. An interesting question is thus whether a similar phenomenon can be observed in a more general setting. For example, in the context of biological machines it is relevant to consider motor traffic on networks as, e.g., the molecular motors involved in intracellular transport move on the cytoskeleton that consists of many interlinked filamentous tracks rather than a single filament \cite{Alberts2007}.

In order to address the aforementioned question we investigate the thermodynamics of motor traffic on Bethe networks subject to hard-core exclusion for two different models of autonomous nanomachines that perform work against a constant load. The first model studied in this paper represents machines where the configurational space of the motor can be projected into an effective one-dimensional, periodic potential energy landscape with a single saddle point or activation barrier, like the potentials studied in \cite{Golubeva2012,VandenBroeck2012}. For high activation barriers, the resulting dynamics is well-described by a network model with a single state corresponding to the minima of the energy landscape. Model I thus describes one-state motors and is solved in sec. \ref{sec:modelI} by adopting the mean-field approach recently introduced in \cite{Neri2011,Neri2013a,Neri2013} for studying exclusion processes on general networks. In sec. \ref{sec:modelIEMP} we optimize the power output with respect to the applied load and find that the corresponding EMP is enhanced, as compared to non-interacting motors with the same parameters, for a wide range of input energies, motor densities and network connectivities due to a phase transition from a heterogeneous to a homogeneous phase as the load is increased. Since the EMP enhancement follows from the effect of load-controlled traffic jams on the resultant motor velocity, it entails a decrease in the maximum power output, as compared to the non-interacting case. We introduce a quantitative measure in order to characterize the trade-off between EMP and the maximum power output when the machines operate at maximum EMP enhancement for fixed network parameters. Using this measure we find that the trade-off is beneficial in a certain region of the network parameter space.

Generally, the configurational space of the motor is more complex than the situation considered in model I. For example, the state space of biological machines contains several intermediate states corresponding to conformational changes of the motor protein and several mechanochemical thermodynamic cycles accounting for multiple dissipative pathways \cite{Nelson2007,Liepelt2007}. Indeed, it has been shown in several works that it is important to include the existence of internal motor states into the description of cooperative effects \cite{Golubeva2012a,Golubeva2013,Nishinari2005,Ciandrini2010,Klumpp2008a}. Furthermore, for loosely-coupled motors, i.e. motors with several thermodynamic cycles, dissipative effects generally lead to a lower efficiency and EMP than for tightly-coupled motors with a single thermodynamic cycle, which the one-state model I is a particular example of \cite{Golubeva2012,Seifert2011a,Golubeva2012a,Golubeva2013}. 
In sec. \ref{sec:modelII} we therefore extend the formalism of \cite{Neri2011,Neri2013a,Neri2013} to study motors with two states and several thermodynamic cycles (model II) in order to explore the robustness of the EMP enhancement to changes in the internal motor dynamics. We find for model II that mutual interactions lead to a boost in the EMP qualitatively similar to the one found for model I whenever the purely dissipative transitions are only moderately strong, which is indeed the case for a broad range of single motor parameters (sec. \ref{sec:modelIIEMP}). Likewise, in this operational regime the EMP enhancement is found to be greater than the corresponding loss in the maximum output power according to the measure described earlier.

\section{Model I: one state}
\label{sec:modelI}
We start out by considering the case of single-state machines moving on periodic Bethe networks comprised of directed segments connected by vertices \cite{Bollobas2001}. The motors act as Poissonian walkers and perform forward (backward) jumps with rate $p$ ($q$) between neighbouring sites on the segment, or between a vertex site and a segment, subject to hard-core exclusion; if a motor attempts a jump to a position already occupied by another motor, the step will be rejected. The dynamics of one-state machines interacting though excluded volume effects is thus equivalent to the paradigmatic model for non-equilibrium transport known as the asymmetric simple exclusion process (ASEP) \cite{ASEP}.

 Thermodynamic consistency requires that the stepping rates satisfy the local detalied balance (LDB) relation $p/q=\exp[\win-\wout]$, where $\win$ and $\wout$ are, respectively, the input work consumed by the motor and the output work delivered by the motor when completing a step in the forward direction. Here and in the following, we measure energies in units of $k_BT$, where $k_B$ is Boltzmanns constant, and $T$ is the temperature. The commonly used parametrisation of the jumping rates based on Kramers theory reads \cite{Kampen1981,Golubeva2012,Seifert2011a,VandenBroeck2012} 
\begin{equation}
\label{eq:rates}
  p=\omega_0 e^{\win-\wout \theta}, \qquad q=\omega_0 e^{\wout (1-\theta)},
\end{equation}
and is thus consistent with the LDB condition. Here, $\omega_0$ is a microscopic jumping rate, and $\theta$ denotes the so-called load factor related to the position of the activation barrier along the reaction coordinate.

The solution for model I builds upon the transport properties of the ASEP on a single open segment, which we therefore briefly revisit here \cite{ASEP}. In this case the particles are injected (removed) with rates $\alpha$ ($\gamma$) at the left end of the segment, and with rates $\delta$ ($\beta$) at the right end. The requirement of current conservation at the left and right segment boundaries allows to define the left and right reservoir densities, $\varrho_l=\varrho[\alpha,\gamma]$ and $\varrho_r=\varrho[-\delta,-\beta]$, respectively, in terms of the injection and removal rates, where
\begin{equation*}
\varrho[x,y]=\frac{p-q+x+y-\sqrt{(p-q+x+y)^2-4(p-q)x}}{2(p-q)}.  
\end{equation*}
The exact phase diagram for the steady-state probability current $j$ in the thermodynamic limit projected onto $\varrho_l$ and $\varrho_r$ consists of three regions termed low-density (LD), high-density (HD), and maximal current (MC) phase, respectively. The current-density relation is given by $j=(p-q)\varrho(1-\varrho)$, where the phase-dependent bulk density $\varrho$ and the phase boundaries are as follows: 
\begin{equation}
\label{eq:rho_asep}
\varrho=
\begin{cases} 
\rhold=\varrho_l & \text{for $1-\varrho_r>\varrho_l$, $\varrho_l<1/2$ (LD)} \\
\rhohd=\varrho_r & \text{for $1-\varrho_r<\varrho_l$, $\varrho_r>1/2$ (HD)} \\
\rhomc=1/2 & \text{for $\varrho_l>1/2, \varrho_r<1/2$ (MC)}.
\end{cases}
\end{equation}

In the case of Bethe networks, all the vertices have identical connectivity and must thus have equal densities which we denote by $\rho_v$. Moreover, all the segment densities must be identical and equal to the total density $\rho$ of motors on the network. For infinitely long homogeneous segments, the vertex and segment occupancies are related through eq. \eqref{eq:rho_asep} and the mean-field entry and exit rates for a segment \cite{Neri2011,Neri2013a,Neri2013},  
\begin{equation}
  \label{eq:effratesI}
  \begin{alignedat}{2}
   \alpha[\rho_v]&=p\rho_v/c,    &\qquad \beta[\rho_v]&=p(1-\rho_v), \\
   \gamma[\rho_v]&=q(1-\rho_v),  &       \delta[\rho_v]&=q\rho_v/c,    
\end{alignedat}
\end{equation}
which only depend on the vertex density $\rho_v$ and the network connectivity $c$. It is worth noting that the entry rates $\alpha$ and $\delta$ are reduced by a factor of $c$ since the particles leaving a vertex enter only one of the $c$ available outgoing segments. It follows from eqs. \eqref{eq:rho_asep}--\eqref{eq:effratesI} that the system undergoes a phase transition from the LD to the HD phase for 
\begin{equation}
 \rhold(\alpha[\rhovb],\gamma[\rhovb])=1-\rhohd(\beta[\rhovb],\delta[\rhovb]) 
\end{equation}
corresponding to the threshold vertex density $\rhovb=c/(c+1)$. All the segments are thus in the LD phase for $\rho<\rhob_l$, where the delimiting segment density $\rhob_l=\rhold[\rhovb]$ is
\begin{equation}
  \rhob_l=\frac{\frac{p+q}{c+1}+(p-q)-\sqrt{(\frac{p+q}{c+1}+(p-q))^2-\frac{4(p-q)p}{c+1}}}{2(p-q)},
\end{equation}
while the high-density conditions apply for $\rho>\rhob_r=\rhohd[\rhovb]=1-\rhob_l$. For intermediate densities a heterogeneous shock phase (SP) arises where a LD zone with density $\rhob_l$ and a HD zone with density $\rhob_r$ coexist on the segment separated by a diffusing domain wall \cite{Neri2013}. In this phase the network responds to increasing densities by growing the HD regions at the expense of the LD regions. The resulting motor current sustained by the network is, however, unaffected since the LD and HD zones have complementary densities $\rhob_r=1-\rhob_l$, and thus carry the same current.      
The current-density profile $J(\rho)$ for model I given by \cite{Neri2013}
\begin{equation*}
\label{eq:Jofrho}
J(\rho)=
\begin{cases}
(p-q)\rho(1-\rho) & \text{for $\rho<\rhob_l$ (LD), $\rhob_r<\rho$ (HD)}  \\
(p-q)\rhob_l(1-\rhob_l) & \text{for $\rhob_l<\rho<\rhob_r$ (SP)}
\end{cases}
\end{equation*}
therefore exhibits a plateau in the shock phase. We note that since $\rhob_{l,r}$ are functions of the jumping rates $p$ and $q$, $J(\rho)$ depends on all the motor parameters through eq. \eqref{eq:rates}.

\begin{figure}
  \centering
  \psfrag{rho}[ct][ct][1.]{$\rho$}
  \psfrag{J}[ct][ct][1.]{$J$}
  \psfrag{LD}[ct][ct][1.]{LD}
  \psfrag{HD}[ct][ct][1.]{HD}
  \psfrag{SP}[ct][ct][1.]{SP}
  \includegraphics[width=\columnwidth]{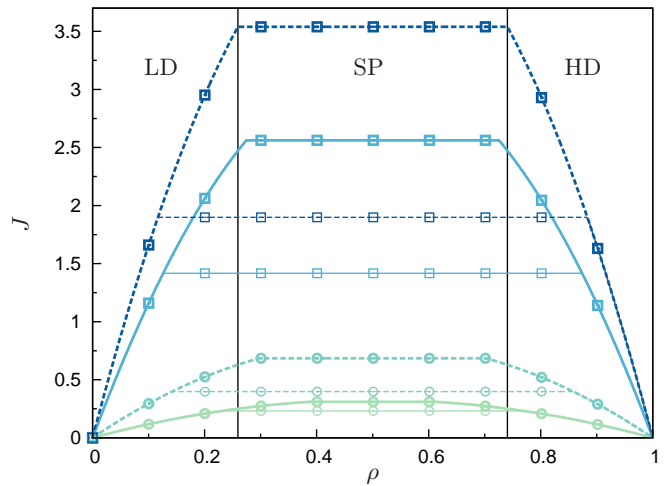}
  \caption{Model I: The current-density relation $J(\rho)$ for two different values of the connectivity, $c=3$ (thick lines) and $c=8$ (thin lines), two different values of the input, $\win=1.5$ (circles) and $\win=3$ (squares), and two different values of the output, $\wout=0.1$ (dashed lines) and $\wout=1$ (solid lines). The vertical lines denote the phase boundaries $\rhob_l$ and $\rhob_r$ for $\win=3$, $\wout=0.1$, $c=3$. Parameters: $\theta=0.3$, $\omega_0=1$.}
  \label{fig:rhoJI}
\end{figure}

In fig. \ref{fig:rhoJI} we show the current-density relation for different values of the input work $\win$, output work $\wout$ and connectivity $c$. It is important to note that the current plateau becomes broader when $c$ and $\win$ are increased, while it shrinks for increasing $\wout$. Hence, for fixed $c$, $\rho$ and $\win$ it is possible to observe a transition from the SP to the LD phase for low segment densities ($\rho<0.5$) or to the HD phase for high segment densities ($\rho>0.5$) as $\wout$ increases. Furthermore, we note that $J(\rho)$ is symmetric around $\rho=0.5$ due to the particle-hole symmetry exhibited by the model.

\subsection{EMP}
\label{sec:modelIEMP}
We proceed by studying the maximum power operation of our model machines. The delivered output power per motor is equal to $\Pout=\wout v$, where $v=J/\rho$ is the average motor velocity. It should be noted that in the SP the velocity is an average over the (domain-averaged) velocities of the motors in the LD and HD domains coexisting on the segment. Similarly, the consumed input power per motor is given by $\Pin=\win r$, where $r$ denotes the average flux of energy input (in units of $\win$). Since the input and output fluxes in model I are tightly coupled by construction, the input rate equals the motor velocity, i.e., $r=v$. The efficiency of the system is thus obtained as $\eta=\Pout/\Pin=\wout/\win$ and is bounded by the value 1, since the extracted work cannot exceed the input work. The upper bound is reached for $p=q$ (see eq. \eqref{eq:rates}) corresponding to reversible conditions under which the power output vanishes, as discussed in the Introduction. We now turn to the question of EMP which is calculated as follows. For given network parameters $c$ and $\rho$ we fix $\win$ and solve $\partial \Pout/\partial \wout |_{\wout*}=0$ for the optimal output $\wout^*$ that maximizes the output power. The EMP is then simply obtained as $\eta^*=\wout^*/\win$, and the above procedure is repeated for increasing values of $\win$. Since the EMP is independent of the microscopic rate $\omega_0$, which only sets a timescale for the motion, we set $\omega_0=1$ throughout this section.
\begin{figure}
 \psfrag{emp}[ct][ct][1.]{$\eta^*$}
 \psfrag{win}[ct][ct][1.]{$\win$}
 \psfrag{legend1}[ct][ct][1.]{$c=5$}
 \psfrag{legend2}[ct][ct][1.]{$c=10$}
 \psfrag{legend3}[ct][ct][1.]{$c=30$}
 \psfrag{legend4}[ct][ct][1.]{single}
 \psfrag{SP}[ct][ct][1.]{{\footnotesize SP}}
 \psfrag{LD}[ct][ct][1.]{{\footnotesize LD}}
 \psfrag{Dmu1}[ct][ct][1.]{{\scriptsize $\win=1$}}
 \psfrag{Dmu2}[ct][ct][1.]{{\scriptsize $\win=2$}}
 \psfrag{Dmu3}[ct][ct][1.]{{\scriptsize $\win=4$}}
 \psfrag{pout}[ct][ct][1.]{$\Pout$}
 \psfrag{wout}[ct][ct][1.]{$\wout$}
 \psfrag{a}[cB][cB][1.]{{\scriptsize $\textbf{(a)}$}}
 \psfrag{b}[cB][cB][1.]{{\scriptsize $\textbf{(b)}$}}
 \psfrag{c}[cB][cB][1.]{{\scriptsize $\textbf{(c)}$}}
 \psfrag{d}[cB][cB][1.]{{\scriptsize $\textbf{(d)}$}}
\centering
\includegraphics[width=\columnwidth]{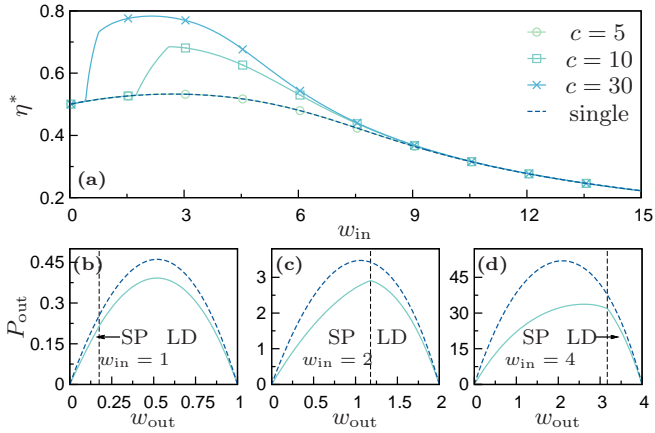}
\caption{Model I: (a) EMP as a function of the input work $\win$ for different values of the network connectivity $c$ (see legend). For comparison, the EMP of non-interacting motors with the same parameter values is plotted with a dashed line. (b)--(d) The output power $\Pout$ as a function of $\wout$ for three different values of $\win$. The solid curve is for the interacting system with $c=10$, and the dashed curve is for the non-interacting case. The vertical dashed line represents the phase boundary between the LD phase and the SP. Parameter values: $\rho=0.15$, $\theta=0.3$.}
   \label{fig:empI_ex}
 \end{figure}
\begin{figure}
 \psfrag{cval}[ct][ct][1.]{$c$}
 \psfrag{rho}[ct][ct][1.]{$\rho$}
  \psfrag{a}[cB][cB][1.]{{\scriptsize $\textbf{(a)}$}}
 \psfrag{b}[cB][cB][1.]{{\scriptsize $\textbf{(b)}$}}
 \psfrag{c}[cB][cB][1.]{{\scriptsize $\textbf{(c)}$}}
\centering
\includegraphics[width=\columnwidth]{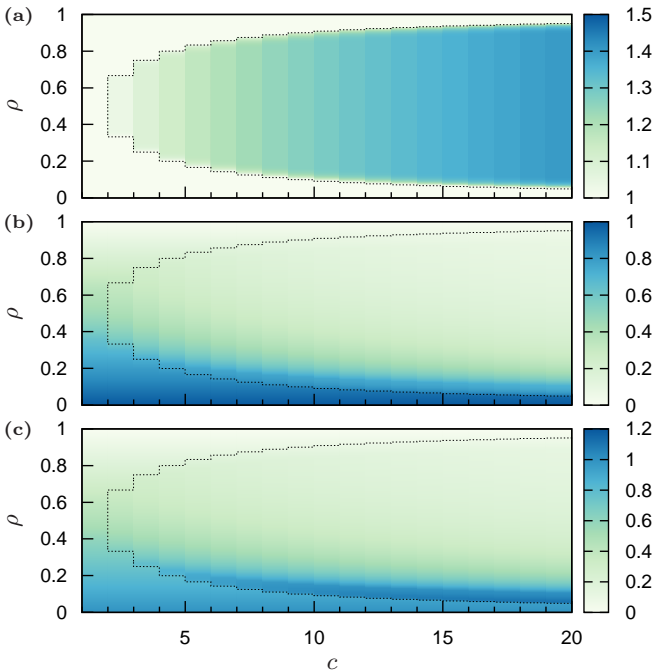}
\caption{Model I: (a) The maximum EMP enhancement $\max_{\win}[\eta^*/\eta^*_0]$ as a function of connectivity $c$ and segment density $\rho$. (b) The ratio $\Pout^{**}/\Poutz^{**}$ of output power for interacting and non-interacting systems at maximum EMP enhancement (see text) as function of $c$ and $\rho$. (c) The product of EMP enhancement $\max_{\win}[\eta^*/\eta^*_0]$ (a) and the ratio of output powers $\Pout^{**}/\Poutz^{**}$ (b) as function of $c$ and $\rho$. In all figures the critical connectivity $\bar{c}(\rho)=\ceil*{1/\rho-1}$ (for $0<\rho<0.5$) is shown with a dotted line, and the load factor is $\theta=0.3$.}
   \label{fig:empI}
 \end{figure}

In fig. \ref{fig:empI_ex}a we consider the behaviour of the EMP as a function of the input work $\win$ for a specific value of the density, $\rho=0.15$, and several values of $c$. For low values of the connectivity, the system is in the LD phase for all values of $\win$ and $\wout$, and the EMP $\eta^*$ is equal to the EMP $\eta^*_0$ obtained for non-interacting motors with the same parameters. This is illustrated by the $c=5$ curve in the figure. However, above a certain critical connectivity $\bar{c}(\rho)$ (see below) the EMP exhibits an enhancement as compared to $\eta_0^*$ for a range of $\win$ values as shown for $c=10$ and $c=30$. The enhancement is caused by a change in the work-velocity relation $v(\wout)$, and hence the output power $\Pout(\wout)$, as illustrated in fig. \ref{fig:empI_ex}b-d for $c=10$. For small values of $\win$ the maximum power output is achieved in the LD phase, and $\eta^*=\eta_0^*$, see fig. \ref{fig:empI_ex}b. As $\win$ increases, the system is in the SP for small output $\wout$, and the machines operate at maximum power at the LD-SP boundary, see fig. \ref{fig:empI_ex}c. The optimal output $\wout^*$ is larger than the corresponding optimal output $\woutz^*$ for single motors, and the EMP $\eta^*=\wout^*/\win$ is hence larger than the single motor EMP $\eta_0^*=\woutz^*/\win$. As $\win$ is increased further, the maximum of the output power always lies in the SP (fig. \ref{fig:empI_ex}d), and $\eta^*>\eta^*_0$. Finally, as $\win\to\infty$, the rates fulfill $q/p \ll 1$ for $\wout \leq \win$, and the densities of the LD and HD zones tend towards $\rhob_l=1/(c+1)$ and $\rhob_r=c/(c+1)$, respectively. Therefore, the SP current $J=(p-q)\rhob_l(1-\rhob_l)$, and hence the velocity $v=J/\rho$, become proportional to the single motor velocity $v_0=p-q$. Thus, $\eta^* \to \eta_0^*$ when $\win$ goes to infinity. 

In order to characterize the dependence of the EMP enhancement on the network parameters, we calculate and plot in fig. \ref{fig:empI}a the maximum EMP enhancement, $\max_{\win}[\eta^*/\eta^*_0]$, for various values of $\rho$ and $c$. As $c$ increases, the maximum enhancement of the EMP becomes larger and occurs at lower values of $\win$ as illustrated in fig. \ref{fig:empI_ex}a. This is due to the fact that the system enters the SP for lower values of $\win$, cf. fig. \ref{fig:rhoJI}. The critical connectivity value $\bar{c}(\rho)$ can be estimated in the following manner for $\rho<0.5$. Since $\rhob_l$ tends asymptotically towards $1/(c+1)$ for $\win\to\infty$, the minimal density required to observe a LD-SP transition for some (large) value of $\win$ is $\rho_{\text{min}}=1/(c+1)$. Since $c$ must be an integer, a good estimate of the critical connectivity is thus given by $\bar{c}(\rho)=\ceil*{1/\rho-1}$, where $\ceil*{\dots}$ denotes the ceiling function. As expected, $\bar{c}(\rho)$ decreases with increasing $\rho$, since the system is closer to the SP as illustrated in fig. \ref{fig:rhoJI}. Due to particle-hole symmetry all the above arguments for the enhancement of the EMP by the existence of the LD-SP transition apply for $1-\rho$ with the LD phase replaced by the HD phase. The maximum EMP enhancement is therefore the same for $\rho$ and $1-\rho$ as can be seen in fig. \ref{fig:empI}a. Finally, it is noteworthy that $\max_{\win}[\eta^*/\eta^*_0]$ only depends weakly on the load factor $\theta$ (data not shown).

In fig. \ref{fig:empI}b we plot the ratio $\Pout^{**}/\Poutz^{**}$ of the output power at maximum enhancement, $\Pout^{**}=\Pout(\win^{**},\wout^*)$, and the corresponding quantity for the non-interacting system, $\Poutz^{**}=\Poutz(\win^{**},\woutz^*)$. Here, $\win^{**}= \argmax_{\win}[\eta^*/\eta^*_0]$ is the input work that optimizes the enhancement, and $\wout^*$ ($\woutz^*$) is the corresponding optimal output work for the interacting (non-interacting) system at $\win^{**}$. The ratio of the output powers is always smaller than one and decreases with increasing density due to exclusion effects. Hence, the EMP enhancement is obtained at the expense of the corresponding maximum power output, as mentioned in the Introduction. 
To study quantitatively the trade-off between EMP enhancement and power loss, we consider in fig. \ref{fig:empI}c the product of the maximum enhacement, $\max_{\win}[\eta^*/\eta^*_0]$, and the ratio $\Pout^{**}/\Poutz^{**}$ of output powers. It is interesting to note that an optimal trade-off between these two quantities is achieved for small values of $\rho$ and values of $c$ slightly higher than $\bar{c}$. In this region of the parameter space the EMP boost caused by mutual exclusion interactions is greater than the corresponding loss in the power output. While this conclusion holds for all values of the load factor, for smaller $\theta$ higher network connectivities and lower densities are required to observe a beneficial trade-off. However, at the same time the overall EMP goes to $1$ as $\theta\to 0$ \cite{Golubeva2012,Seifert2011a,VandenBroeck2012}, which offers a possibility for switching between high EMP and good EMP-maximum power trade-off regime by tuning the load factor in a fixed network setup.  
Finally, it is worth noting that a qualitatively similar picture is obtained if instead the quantity $\max_{\win}[\eta/\eta_0\, \Pout/\Poutz]$ is used as a measure of the EMP-maximum power trade-off; this quantity reaches values larger than 1 in approximately the same region of the ($c$,$\rho$)-space as the quantity considered in fig. \ref{fig:empI}c.  
 
\section{Model II: two states}
\label{sec:modelII}
\begin{figure}
  \centering
  \includegraphics[width=.9\columnwidth]{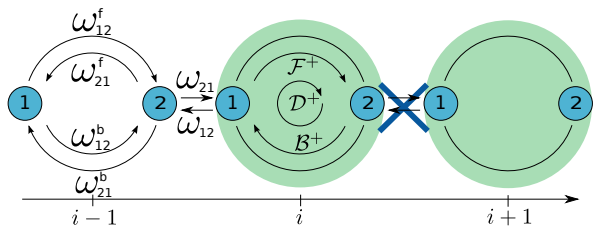}
  \caption{Model II: The two-state model contains a forward stepping dicycle $\mathcal{F}^+$, a backward stepping dicycle $\mathcal{B}^+$ and a dissipative dicycle $\mathcal{D}^+$ together with their reverse dicycles (not shown). The stepping transitions characterised by the jumping rates $\omega_{21}$ and $\omega_{12}$ are subject to the ASEP exclusion rule and are therefore forbidden if the target site is already occupied, as illustrated in the cartoon for sites $i$ and $i+1$.}
  \label{fig:modelII}
\end{figure}
In order to explore the effect of internal dissipation on the EMP we proceed by studying the two-state model depicted in fig. \ref{fig:modelII}. Inspired by the mechanochemical operation of two-headed molecular motors \cite{Liepelt2007} we assume that the machine contains two symmetric stepping cycles. The dicycle $\mathcal{F}^+$ consisting of the transitions with rates $\omega_{12}^f$ and $\omega_{21}$, see fig. \ref{fig:modelII}, utilises the input work $\win$ to perform a forward mechanical step, thereby generating the output work $\wout$. Similarly, the dicycle $\mathcal{B}^+$ made up of transitions with rates $\omega_{21}^b$ and $\omega_{12}$ represents a backward mechanical step with input work $\win$ and delivered work $-\wout$. The model also contains a purely dissipative dicycle $\mathcal{D}^+$ consisting of $\omega_{12}^f$ followed by $\omega_{21}^b$ in which the motor consumes the input work $2\win$ and performs no output work. The dicycles describing the above-mentioned processes operating in reverse are denoted by $\mathcal{F}^-$, $\mathcal{B}^-$ and $\mathcal{D}^-$, respectively. The excluded volume effects affect the transitions characterised by the rates $\omega_{12}$ and $\omega_{21}$, since these transitions represent mechanical steps between neighbouring lattice sites.

The transition rates of the model can be written as
\begin{equation}
  \label{eq:ratesII}
  \begin{alignedat}{2}
    \omega_{21}&=\omega_{21,0}e^{-\wout \theta} &\quad  \omega_{12}^f&=e^{\win}\omega_{21}^f \omega_{12,0}/\omega_{21,0} \\
     \omega_{12}&=\omega_{12,0}e^{\wout(1-\theta)}  &  \omega_{21}^b&=e^{\win}\omega_{12}^b \omega_{21,0}/\omega_{12,0},         
\end{alignedat}
\end{equation}
where the expressions for $\omega_{12}^f$ and $\omega_{21}^b$ follow from the LDB relations for the forward and backward cycles,
\begin{equation}
  \label{eq:DB}
  e^{\win}=\frac{\omega_{12}^f \omega_{21,0}}{\omega_{21}^f \omega_{12,0}}=\frac{\omega_{21}^b \omega_{12,0}}{\omega_{12}^b \omega_{21,0}}.
\end{equation}
Since the forward and backward steps are assumed to be triggered by identical processes, we take $\omega_{12}^f=\omega_{21}^b$, which in turn leads to $\omega_{21}^f=\omega_{12}^b \left( \omega_{21,0}/\omega_{12,0}\right)^2$. After fixing the timescale by setting $\omega_{12,0}=1$ we are then left with three model parameters, namely $\omega_{21,0}$, $\omega_{12}^b$ and $\theta$. The tight coupling condition for non-interacting motors is acquired by minimising the probability current carried by the backward cycle $\mathcal{B}$ relative to the mechanical current (or velocity). Far from equilibrium, $\exp[\win-\wout]\gg 1$, such a condition for tight coupling can be written as
\begin{equation}
  \label{eq:tc}
  \omega_{12}^b \ll  [1-e^{\wout/\omega_{21}^0}]e^{-(\win+\theta\wout)}.
\end{equation}
 We can thus control the extent of the tight coupling regime in the low-density limit by changing the magnitudes of $\omega_{12}^b$ and $\omega_{21,0}$.

In analogy to model I, the solution for model II is based on the mean-field phase diagram of the corresponding problem on an infinitely long open segment. The latter is obtained using the maximal current principle \cite{MCH} previously described in detail in \cite{Golubeva2012a,Golubeva2013}. Since the transitions $\omega_{12,21}^b$ of the backward cycle reduce the motor velocity while contributing to the total consumed energy, the mechanical output current $\jsm(\varrho)$ and the input current $\jsc(\varrho)$ as a function of density for a single segment can be expressed as
\begin{align}
   \jsm=&j(\mathcal{F})-j(\mathcal{B})=(\omega_{12}^f\varrho^1-\omega_{21}^f\varrho^2)-(\omega_{21}^b\varrho^2-\omega_{12}^b\varrho^1) \nonumber \\
   \jsc=&j(\mathcal{F})+j(\mathcal{B})=(\omega_{12}^f\varrho^1-\omega_{21}^f\varrho^2)+(\omega_{21}^b\varrho^2-\omega_{12}^b\varrho^1), \nonumber
\end{align}
respectively. Here, $j(\mathcal{C})=j(\mathcal{C}^+)-j(\mathcal{C}^-)$ with $\mathcal{C}=\mathcal{F},\mathcal{B}$ is the probability current carried by the thermodynamic cycle $\mathcal{C}$, and the populations $\varrho^1$ and $\varrho^2$ of the two motor states obey the master equation 
\begin{equation}
\label{eq:ME}
  0=(\omega_{21}^f+\omega_{21}^b)\varrho^2-(\omega_{12}^f+\omega_{12}^b)\varrho^1+(\omega_{21}\varrho^2-\omega_{12}\varrho^1)(1-\rho)
\end{equation}
with the normalisation condition $\varrho^1+\varrho^2=\varrho$. The densities of the left and right reservoirs, $\varrho_l$ and $\varrho_r$, respectively, can be found by solving eq. \eqref{eq:ME} together with the current conservation conditions at the segment boundaries,
\begin{equation*}
 \jsm(\varrho_l)=\alpha(1-\varrho_l)-\gamma \varrho^1_l  \text{  and  } \jsm(\varrho_r)=\beta \varrho_r^2+\delta(1-\varrho_r).
\end{equation*}
The maximal current principle can be thought of as a variational statement for the bulk density $\varrho$ based on an optimization of the mechanical current $\jsm$ depending on the relative values of the reservoir densities $\varrho_l$ and $\varrho_r$ \cite{MCH},
\begin{equation}
  \label{eq:mch}
  \varrho=
\begin{cases}
\argmax \limits_{\tilde{\varrho}\in [\varrho_r,\varrho_l]} \jsm(\tilde{\varrho}) & \text{for $\varrho_l>\varrho_r$} \\
\argmin \limits_{\tilde{\varrho}\in [\varrho_l,\varrho_r]} \jsm(\tilde{\varrho}) & \text{for $\varrho_l<\varrho_r$}. 
\end{cases}
\end{equation}
The phase diagram predicted by eq. \eqref{eq:mch} consists of three phases as introduced in model I, where the densities and the corresponding phase boundaries now are given by
\begin{equation*}
\label{eq:rhoII}
\varrho= \begin{cases}
\rhold=\varrho_l[\alpha,\gamma] & \text{for $\varrho_l<\varrho^*$, $\jsm(\varrho_l)<\jsm(\varrho_r)$} \\
\rhohd=\varrho_r[\beta,\delta] & \text{for $\varrho_r >\varrho^*$, $\jsm(\varrho_l)>\jsm(\varrho_r)$} \\
\rhomc=\varrho^* & \text{for $\varrho_l>\varrho^*$, $\varrho_r <\varrho^*$}
\end{cases} \nonumber
\end{equation*}
with $\varrho^*=\max_{\varrho}\jsm(\varrho)$. 

By considering the probability currents between the segment ends and the corresponding vertex sites, the effective boundary rates for a homogeneous segment in a Bethe network are found to be
\begin{equation}
\label{eq:effratesII}
\begin{alignedat}{2}
  \alpha[\rho_v^2]&=\omega_{21} \rho_v^2/c,     &\qquad \beta[\rho_v]&=\omega_{21}(1-\rho_v), \\
  \gamma[\rho_v]&=\omega_{12}(1-\rho_v),      &       \delta[\rho_v^1]&=\omega_{12} \rho_v^1/c, 
\end{alignedat}
\end{equation}
where $\rho_v^i$ denotes the density of motors in state $i$ at a vertex site. In analogy to model I, the segment densities $\rho^i$ and the vertex densities $\rho_v^i$ are identical for all the segments and vertices. The relevant quantity to consider is therefore the total threshold vertex density $\rhovb$ and its components $\rhovb^i$, which are obtained by solving the equation defining the LD-HD phase boundary, $\jsm(\rhold[\rhovb^i])=\jsm(\rhohd[\rhovb^i])$, together with a condition for current conservation at the vertex,
\begin{align}
   0=&c  \omega_{21}\varrho_r^2[\rhovb^i](1-\rhovb)+(\omega_{21}^f+\omega_{21}^b)\rhovb^2 \nonumber\\
    &-\rhovb^1 \omega_{12}(1-\varrho_r[\rhovb^i])-(\omega_{12}^f+\omega_{12}^b)\rhovb^1,  \label{eq:pv1}
\end{align}
and the requirement that $\rhovb=\rhovb^1+\rhovb^2$. Here, e.g., $\rhold[\rhovb^i]$ is short-hand notation for $\rhold$ as a function of $\rhovb^1$, $\rhovb^2$ and $\rhovb$ through the effective rates, eq. \eqref{eq:effratesII}. We note that the critical value $\rhovb$ now depends on the connectivity $c$ and on all the transition rates, as opposed to model I where $\rhovb$ was only a function of $c$. 

\begin{figure}
  \centering
  \psfrag{rho}[ct][ct]{$\rho$}
  \psfrag{J}[ct][ct]{$\jm$, $\jc$}
  \psfrag{ylabelvar}[ct][ct]{$\jm/\jc$}
  \psfrag{rho}[ct][ct]{$\rho$}
  \psfrag{r1}[cc][cc]{\scriptsize $\omega_{21,0}$}
  \psfrag{r2}[cc][cc]{\scriptsize $\omega_{12}^b$}
  \psfrag{rho}[ct][ct]{$\rho$}
  \psfrag{legendlegend1}[ct][ct]{\scriptsize $10^4 \quad 10^{-4}$}
  \psfrag{legendlegend2}[ct][ct]{\scriptsize $10^3 \quad 10^{-4}$}
  \psfrag{legendlegend3}[ct][ct]{\scriptsize $10^2 \quad 10^{-4}$}
  \psfrag{legendlegend4}[ct][ct]{\scriptsize $10^4 \quad 10^{-3}$}
  \psfrag{legendlegend5}[ct][ct]{\scriptsize $10^4 \quad 10^{-2}$}
  \psfrag{a}[cB][cB][1.]{{\scriptsize $\textbf{(a)}$}}
  \psfrag{b}[cB][cB][1.]{{\scriptsize $\textbf{(b)}$}}
  \includegraphics[width=\columnwidth]{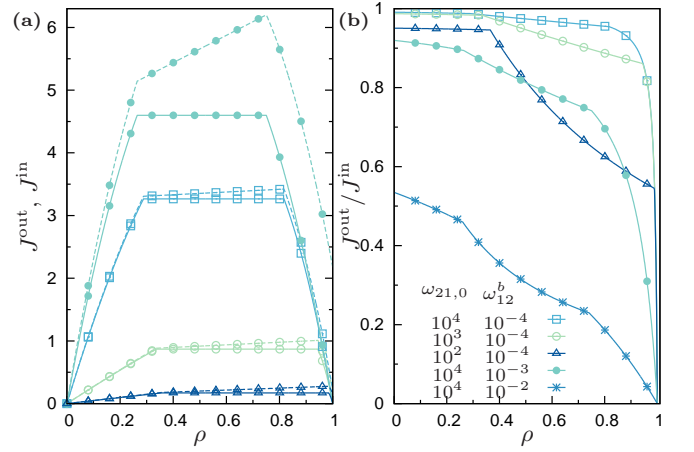}
  \caption{Model II: (a) The mechanical current $\jm$ (solid lines) and the chemical current $\jc$ (dashed lines) as function of the density $\rho$ for different values of the transition rates $\omega_{21,0}$ and $\omega_{12}^b$ as specified in the legend in (b). Note that the values for $\omega_{21,0}=10^4$ and $\omega_{12}^b=10^{-2}$ are of a different order of magnitude and therefore not shown in the figure. (b) The coupling ratio $v/r=\jm/\jc$ of the mechanical and chemical coordinates as function of $\rho$. Parameters: $\win=3.5$, $\wout=0.7$, $\theta=0.3$, $c=3$, $\omega_{12,0}=1$, and the rest of the transition rates are calculated as described in the text.}
  \label{fig:rhoJII}
\end{figure}

In the SP delimited from the left and right by the densities $\rhob_l=\rhold[\rhovb^i]$ and $\rhob_r=\rhohd[\rhovb^i]$, respectively, the mechanical output current $\jm(\rho)$ exhibits a plateau with the constant value $\jm(\rho)=\jsm(\rhob_l)=\jsm(\rhob_r)$. The input current $\jc(\rho)$ is, however, a linear interpolation between the LD value $\jsc(\rhob_l)$ and the HD value $\jsc(\rhob_r)$, since the fraction of low-density sites increases linearly with $\rho$ in the SP. In the LD ($\rho<\rhob_l$) and HD ($\rho>\rhob_r$) phases the output and input currents are given by $\jm(\rho)=\jsm(\varrho)$ and $\jc(\rho)=\jsc(\varrho)$, respectively. In fig. \ref{fig:rhoJII}a we plot $\jm(\rho)$ and $\jc(\rho)$ for different values of the transition rates $\omega_{21,0}$ and $\omega_{12}^b$ for fixed values of $\win$, $\wout$ and $c$. We note that the introduction of internal states breaks the particle-hole symmetry, i.e. $J^{\text{in,out}}(\rho)\neq J^{\text{in,out}}(1-\rho)$, even in the tightly coupled limit $\jm/\jc = 1$ (see discussion below). The dependence of the current-density relations on the values of $\win$, $\wout$ and $c$ in model II is qualitatively similar to the one observed for model I, cf. fig. \ref{fig:rhoJI}, and is therefore not shown in fig. \ref{fig:rhoJII}a.

Fig. \ref{fig:rhoJII}b shows the coupling ratio $v/r=\jm/\jc$, where $v=\jm/\rho$ is the average velocity, and $r=\jc/\rho$ denotes the average energy input rate. In general, in the limit $\rho\to 0$ the coupling ratio is equal to the corresponding ratio for non-interacting motors, while it decreases with increasing $\rho$ as the effect of the steric interactions becomes more pronounced. Finally, the ratio goes to $0$ as $\rho\to 1$, since the velocity vanishes in this limit, while the energy input rate remains finite. For $\omega_{21,0}=10^4$ and $\omega_{12}^b=10^{-4}$ the coupling ratio is close to $1$ for a wide range of densities, where $\jm/\jc=1$ defines the tightly coupled limit. For other parameter values the coupling ratio deviates more strongly from the tight-coupling behaviour in accordance with eq. \eqref{eq:tc}.

\subsection{EMP}
\label{sec:modelIIEMP}
We proceed by calculating the EMP for model II using the same procedure as for model I. However, since in general the motors are loosely coupled, $v\neq r$, and the EMP is given by $\eta^*=\wout^*v^*/(\win r^*)$, where $v^*=v(\wout^*)$ and $r^*=r(\wout^*)$ denote the input and output rates evaluated at the optimal load $\wout^*$.     

\begin{figure}
  \centering
  \psfrag{win}[ct][ct]{$\win$}
  \psfrag{emp}[ct][ct]{$\eta^*$}
  \psfrag{legendlegend1}[ct][ct]{\scriptsize $\rho=0.7$, $c=8$}
  \psfrag{legendlegend2}[ct][ct]{\scriptsize $\rho=0.7$, $c=3$}
  \psfrag{legendlegend3}[ct][ct]{\scriptsize $\rho=0.3$, $c=8$}
  \psfrag{legendlegend4}[ct][ct]{\scriptsize $\rho=0.3$, $c=3$}
  \psfrag{a}[lB][lB][1.]{{\scriptsize $\textbf{(a)}$  $\omega_{21,0}=10^4$, $\omega_{12}^b=10^{-4}$}}
  \psfrag{c}[lB][lB][1.]{{\scriptsize $\textbf{(c)}$ $\omega_{21,0}=10^4$, $\omega_{12}^b=10^{-3}$}}
  \psfrag{d}[lB][lB][1.]{{\scriptsize $\textbf{(d)}$ $\omega_{21,0}=10^2$, $\omega_{12}^b=10^{-4}$}}
  \psfrag{b}[lB][lB][1.]{{\scriptsize $\textbf{(b)}$ $\omega_{21,0}=10^3$, $\omega_{12}^b=10^{-4}$}}
  \includegraphics[width=\columnwidth]{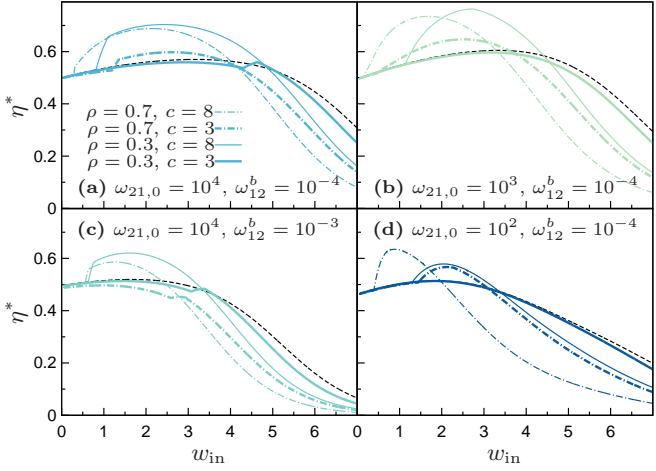}
  \caption{Model II: The efficiency at maximum power $\eta^*$ as a function of the input $\win$ for different values of the density $\rho$ and connectivity $c$ as specified in the legend in (a), and for different transition rates $\omega_{21,0}$ and $\omega_{12}^b$ as specified in the subfigures. In each subfigure the corresponding EMP for non-interacting motors with the same parameters is plotted with a dashed line. Parameters that are not specified in the legends are the same as in fig. \ref{fig:rhoJII}.}
  \label{fig:empII}
\end{figure}
In fig. \ref{fig:empII} we plot the EMP as a function of $\win$ for different values of the transition rates $\omega_{21,0}$ and $\omega_{12}^b$ and for different values of $\rho$ and $c$. The curves for $\omega_{21,0}=10^4$ and $\omega_{12}^b=10^{-4}$ shown in fig. \ref{fig:empII}a represent the most tightly coupled system that we consider here. The absence of particle-hole symmetry in model II leads to different behaviours of the EMP as a function of $\win$ for densities $\rho$ and $1-\rho$, as can be seen in the figure for the case $\rho=0.3$ and $\rho=0.7$. Since the SP is shifted to higher densities, see fig. \ref{fig:rhoJII}a, the enhancement of the EMP for $\rho=0.3$ generally occurs at higher values of $\win$ as compared to $1-\rho=0.7$. However, for larger connectivities the current-density relation becomes more symmetric in $\rho$, and the EMP enhancement thus takes place at similar $\win$ for $\rho$ and $1-\rho$. It is important to note that the presence of additional futile energy dissipation caused by motor-motor interactions generally decreases the EMP and the enhancement of the EMP at large values of $\rho$. As discussed in the previous section, the extent of traffic jam induced futile energy dissipation depends on the transition rates, see fig. \ref{fig:rhoJII}b, and increases with increasing $\win$ and $\rho$. For the values of $\omega_{21,0}$ and $\omega_{12}^b$ used in fig. \ref{fig:empII}a the coupling ratio varies little with density for small $\win$, and the EMP enhancement for $c=8$ is thus similar for the two values of $\rho$ for $\win \lesssim 3$. 

\begin{figure}
  \centering
  \psfrag{pout}[ct][ct]{$\Pout$}
  \psfrag{eta}[ct][ct]{$\eta$}
  \psfrag{wout}[ct][ct]{$\wout$}
  \psfrag{Ia}[lB][lB]{{\scriptsize $\textbf{(Ia)}$}}
  \psfrag{Ib}[lB][lB]{{\scriptsize $\textbf{(Ib)}$}}
  \psfrag{IIa}[lB][lB]{{\scriptsize $\textbf{(IIa)}$ $\rho=0.3$}}
  \psfrag{IIb}[lB][lB]{{\scriptsize $\textbf{(IIb)}$ $\rho=0.7$}}
  \psfrag{SP}[ct][ct]{{\footnotesize SP}}
  \psfrag{LD}[ct][ct]{{\footnotesize LD}}
  \psfrag{HD}[ct][ct]{{\footnotesize HD}}
  \psfrag{n}[cc][cc]{{\footnotesize $\eta^*$}}
  \psfrag{n0}[cc][cc]{{\footnotesize $\eta^*_0$}}
  \includegraphics[width=\columnwidth]{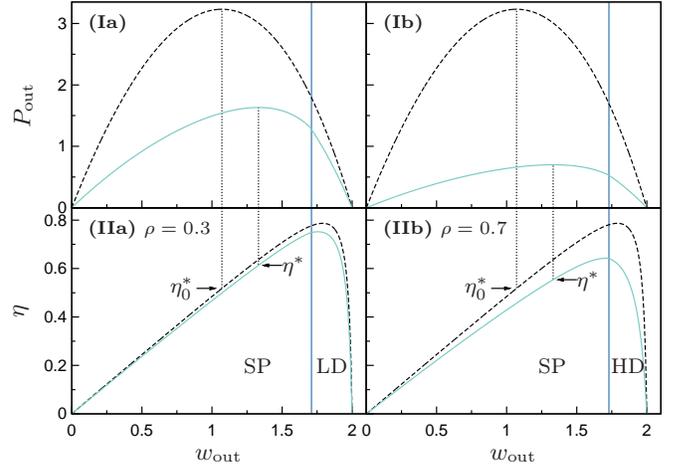}
  \caption{Model II: The output power $\Pout$ (row I) and efficiency $\eta$ (row II) as function of the output $\wout$ for two different values of the density, $\rho=0.3$ (a) and $\rho=0.7$ (b). The curves for the interacting and non-interacting system are shown with solid and dashed lines, respectively. The solid vertical lines denote the LD-SP and HD-SP phase transitions for $\rho=0.3$ and $\rho=0.7$, respectively. The dotted vertical lines indicate the position of maximum output power and EMP. The figure illustrates that the EMP $\eta^*$ is reduced for $\rho=0.7$ as compared to $\rho=0.3$ due to futile energy dissipation. Parameter values are $\omega_{21,0}=10^4$, $\omega_{12}^b=10^{-3}$, $c=3$ and $\win=2$.}
  \label{fig:poutII}
\end{figure}

Fig. \ref{fig:empII}c shows the behaviour of the EMP for $\omega_{21,0}=10^4$ and $\omega_{12}^b=10^{-3}$. For these parameter values the particle-hole symmetry is approximately restored as can be seen in fig. \ref{fig:rhoJII}a. As a consequence, the EMP enhancement region is located at approximately the same $\win$ values for $\rho$ and $1-\rho$. However, the coupling ratio now depends more strongly on $\rho$ than in \ref{fig:empII}a, and the EMP is thus lowered significantly for $\rho=0.7$, with the strongest suppression occuring for large $\win$. The effect of futile combustion on the EMP enhancement is illustrated in fig. \ref{fig:poutII} for $\win=2$ and $c=3$. The optimal force $\wout^*$ maximizing the power output is the same for $\rho=0.3$ and $\rho=0.7$ (row I), since the mechanical output current in the SP is independent of density. However, the incoming energy current is larger for $\rho=0.7$ due to dissipation, and the resulting EMP is thus lower. 

As the value of $\omega_{21,0}$ is decreased, the current-density relation becomes more asymmetric. As a result, for $\rho=0.3$ and $c=8$ the enhancement region lies at slightly higher values of $\win$ (fig. \ref{fig:empII}b,d). For $\rho=0.3$ and $c=3$ the transition to the SP takes place at large values of $\win$ where, in anology to model I, the velocity becomes proportional to the single-motor velocity. Hence, the EMP enhancement does not occur for these values of $\rho$ and $c$.   
For $\rho=0.7$ and $\omega_{21,0}=10^3$ (fig. \ref{fig:empII}b) the EMP enhancement region is shifted to lower $\win$ values as compared to fig. \ref{fig:empII}a since it is caused by a LD-SP transition rather than a HD-SP transition as in \ref{fig:empII}a. When $\omega_{21,0}$ is lowered further to $10^2$ (fig. \ref{fig:empII}d), the enhancement region moves back to higher $\win$ values, since the increased particle-hole asymmetry has placed the system further away from the SP transition in this case as compared to \ref{fig:empII}b. Furthermore, the EMP is decreased for $\rho=0.7$ as compared to \ref{fig:empII}b due to additional internal dissipation.

\begin{figure}
  \centering
  \psfrag{a}[lB][lB][1.]{\scriptsize $\textbf{(a)}$}
  \psfrag{c}[lB][lB][1.]{\scriptsize $\textbf{(c)}$}
  \psfrag{d}[lB][lB][1.]{\scriptsize $\textbf{(d)}$}
  \psfrag{b}[lB][lB][1.]{\scriptsize $\textbf{(b)}$}
  \psfrag{cval}[ct][ct]{$c$}
  \psfrag{rho}[ct][ct]{$\rho$}
  \includegraphics[width=\columnwidth]{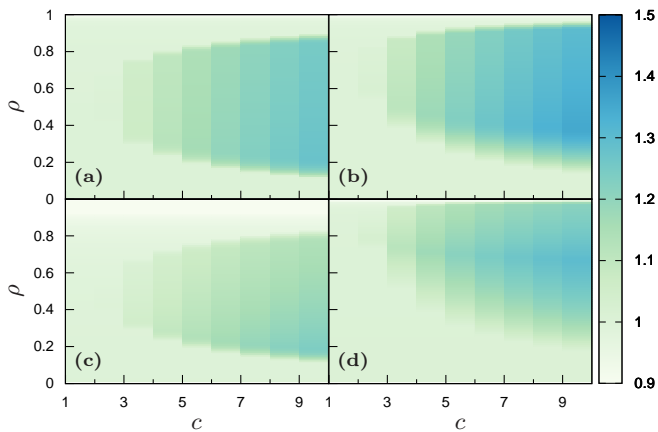}
  \caption{Model II: The maximum EMP enhancement $\max_{\win}[\eta^*/\eta^*_0]$ as a function of connectivity $c$ and segment density $\rho$. The parameter values used are identical to those used in the corresponding subfigures of fig. \ref{fig:empII}.}
  \label{fig:empIIall}
\end{figure}
In fig. \ref{fig:empIIall} we show the dependence of the EMP enhancement optimized with respect to the input work, $\max_{\win}[\eta^*/\eta^*_0]$, on the network connectivity and motor density for the single-motor parameters employed in fig. \ref{fig:empII}. In analogy to model I we find that this quantity is larger than 1 above a certain connectivity threshold which, however, now is a complicated function of all the model parameters. From fig. \ref{fig:empIIall}b,d it is evident that for small densities the breaking of particle-hole symmetry in model II pushes the critical connectivity to higher values as compared to fig. \ref{fig:empI}a, while for high densities the EMP boost occurs at smaller network connectivities than observed for model I. When the flux carried by the dissipative cycle $\mathcal{D}$ becomes significantly high, the EMP and, hence, the EMP enhancement are suppressed with the high motor density regime being affected the most as illustrated in fig. \ref{fig:empIIall}c. It should be noted that the EMP is also enhanced by interactions for more loosely coupled machines than the examples considered in this section. However, in these cases the EMP boost only occurs at unrealistically high network connectivities while the absolute value of the EMP remains low, and such machines are therefore of limited practical interest.

\begin{figure}
  \centering
  \psfrag{a}[lB][lB][1.]{\scriptsize $\textbf{(a)}$}
  \psfrag{c}[lB][lB][1.]{\scriptsize $\textbf{(c)}$}
  \psfrag{d}[lB][lB][1.]{\scriptsize $\textbf{(d)}$}
  \psfrag{b}[lB][lB][1.]{\scriptsize $\textbf{(b)}$}
  \psfrag{cval}[ct][ct]{$c$}
  \psfrag{rho}[ct][ct]{$\rho$}
  \includegraphics[width=\columnwidth]{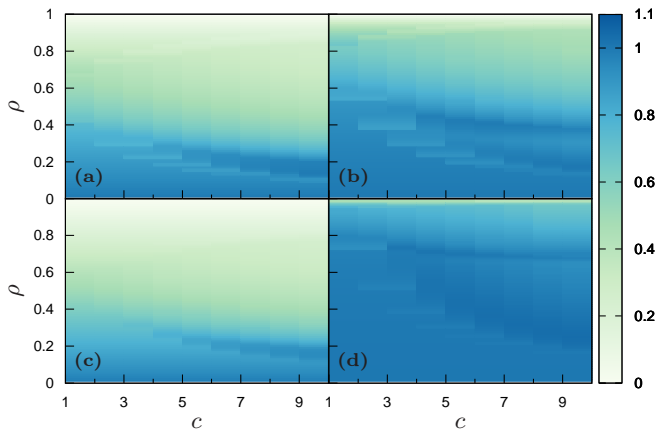}
  \caption{Model II: The product of the maximum EMP enhancement $\max_{\win}[\eta^*/\eta^*_0]$, fig. \ref{fig:empII}, and the ratio of output powers $\Pout^{**}/\Poutz^{**}$ corresponding to fig. \ref{fig:empIIall} as a function of $c$ and $\rho$.}
  \label{fig:emptimespoutII}
\end{figure}
The competition between the EMP enhancement due to mutual interactions and the accompanying loss of maximum power output when compared to individual motors is considered in fig. \ref{fig:emptimespoutII} using the measure introduced previosly in sec. \ref{sec:modelIEMP}. The product of the maximum EMP enhancement $\max_{\win}[\eta^*/\eta^*_0]$ and the corresponding ratio of output powers at maximum EMP enhancement, $\Pout^{**}/\Poutz^{**}$, plotted in fig. \ref{fig:emptimespoutII} exhibits a behaviour qualitatively similar to the one observed for model I in fig. \ref{fig:empI}c. Interestingly, a strongly asymmetric current-density relation entails that the quantity measuring the trade-off is approximately $1$ for all motor densities (fig. \ref{fig:emptimespoutII}d), since the velocity, and therefore the power, decrease most quickly with density in the HD phase, which is in turn only entered for $\rho\sim 1$. Hence, in this case we find the somewhat counter-intuitive result that systems operating at optimal EMP enhancement at high motor densities only experience an insignificant decrease in the power output due to mutual exclusion.

In summary, the maximum power operation of loosely coupled, interacting machines moving on a network is a result of an intricate interplay between the asymmetric mechanical current-density relation and interaction-induced futile combustion, which is governed by the network parameters $c$ and $\rho$ as well as the values of the single-motor transition rates through the LDB constraints \eqref{eq:DB}. However, for the parameters considered in figs. \ref{fig:empII}--\ref{fig:emptimespoutII} which represent moderate deviations from the tightly-coupled behavior we find that all the conclusions of sec. \ref{sec:modelIEMP} found in the context of the simpler model I remain valid in the presence of internal motor dynamics. 

\section{Conclusions} 
We have studied the EMP of autonomous motors operating on a Bethe network under a constant load for tightly coupled one-state motors (model I) and loosely coupled two-state motors with several thermodynamic cycles (model II). For both models we find that, above a certain density-dependent critical network connectivity, mutual exclusion interactions enhance the EMP due to an altered response of the system to externally applied loads. Furthermore, by considering the product of 1) the EMP enhancement maximized with respect to the input work and 2) the corresponding ratio of output powers for interacting and non-interacting motors, we find for a range of network connectivities and motor densities that the EMP enhancement compensates for the loss in the output power induced by exclusion. We have investigated the robustness of such a beneficial trade-off to changes in the internal motor dynamics and provided some strategies for designing motor traffic systems that operate efficiently at maximum power without a significant decrease in the maximum power due to jamming effects.  

As mentioned in the Introduction, the present work was inspired by the observation that collective motor traffic of kinesin motors on a single filament exhibits an interaction-induced EMP enhancement for a variety of different boundary conditions and model parameter values \cite{Golubeva2012a,Golubeva2013}. In this paper we have shown that a qualitatively similar behaviour can be observed in the more general context of exclusion processes on Bethe networks. Furthermore, we also expect our conclusions to apply for heterogeneous networks such as, e.g., Poissonian networks \cite{Bollobas2001} where the relative number of low-density and high-density segments would change with the applied load \cite{Neri2011,Neri2013a,Neri2013}, thereby causing a similar response of the velocity to external load as the one found for homogeneous networks. We therefore believe that our findings are relevant more generally for many-motor systems with an altered characteristic response to external driving as a consequence of mutual interactions.

\acknowledgements
The authors gratefully acknowledge financial support from Lundbeck Fonden and the Danish Council for Independent Research.

%

\end{document}